%% file: main.tex
\documentclass{fundam}
\usepackage{lcbs-macros}
\hypersetup{pdfauthor={Md. Tanzeem Rahat, Md. Manzurul Hasan, Sadia Sharmin},pdftitle={The Longest Common Bitonic Subsequence: Match-Sensitive Algorithms and Conditional Hardness}}

\begin{document}
\title{The Longest Common Bitonic Subsequence: Match-Sensitive Algorithms and Conditional Hardness}
\author{Md. Tanzeem Rahat\textsuperscript{1}\\
\textup{\textbf{Md. Manzurul Hasan\textsuperscript{1}\corresponding, Sadia Sharmin\textsuperscript{2}}}\\
\textsuperscript{1}Department of Computer Science\\
American International University-Bangladesh, Dhaka, Bangladesh\\
\textsuperscript{2}Department of Computer Science and Engineering\\
Bangladesh University of Engineering and Technology, Dhaka 1000, Bangladesh\\
\texttt{tanzeem.rahat@aiub.edu}\\
\texttt{manzurul@aiub.edu}, \texttt{sadiasharmin@cse.buet.ac.bd}}
\maketitle
\runninghead{M. T. Rahat, M. M. Hasan, S. Sharmin}{Longest Common Bitonic Subsequence}

\begin{abstract}
The longest common bitonic subsequence problem asks for a longest common
subsequence of two ordered sequences whose values strictly increase and then
strictly decrease; either phase may be empty. We formulate the problem through
increasing and decreasing endpoint values at matching position pairs. This
gives a constructive quadratic baseline and a match-sensitive algorithm based
on two standard dominance-maximum passes. Its time is the sum of an
input-sorting term and the number of matches times a squared logarithmic
factor. We state the endpoint interface that permits reuse of increasing
subsequence algorithms, and distinguish this specialization from new range
searching machinery. A linear-size padding reduction transfers the
conditional strongly subquadratic lower bound for longest common increasing
subsequence to the bitonic problem. Reproducible implementations, exhaustive
small-instance checks, and newly measured synthetic experiments document
correctness and the practical tradeoff between sparse and dense processing.
\end{abstract}
\noindent\textbf{ACM subject classification:} Theory of computation:
Design and analysis of algorithms; Dynamic programming; Problems, reductions
and completeness.

\section{Introduction}
For two sequences $A[1..n]$ and $B[1..m]$ over a totally ordered alphabet,
a common subsequence must respect the position order in both inputs. The
longest common bitonic subsequence (LCBS) additionally requires a strict
increase to one peak, followed by a strict decrease. Purely increasing and
purely decreasing sequences are allowed. We write $\LCBS(A,B)$ for the optimum
length, with value zero when the inputs have no common symbol.

The problem combines a simple value-shape constraint with two embedding
orders. Let $V=\{(i,j):A[i]=B[j]\}$ and $M=|V|$. For a match $v$, let $I(v)$
be the longest increasing common subsequence ending exactly at $v$, and let
$D(v)$ be the longest decreasing common subsequence starting exactly at $v$.
The central identity is
\begin{equation}\label{eq:peak}
 \LCBS(A,B)=\max_{v\in V}\{I(v)+D(v)-1\},
\end{equation}
with the empty maximum interpreted as zero. Anchoring both halves at the same
match makes their embeddings compatible; the peak is counted only once.

Our main algorithm evaluates these endpoint arrays using two sweeps over the
matches and standard two-dimensional dominance maxima. Put $S=n+m$ and
$\lg_+ x=\log_2(2+x)$. The resulting time bound is
\begin{equation}\label{eq:bound}
 O\bigl(S\lg_+S+M\lg_+^2M\bigr),
\end{equation}
and total auxiliary space is $O(S+M\lg_+M)$, including preprocessing.
The dynamic-programming structure alone occupies
$O(M\lg_+M)$ space. These bounds include recovery of an optimal subsequence.
The distinction between total workspace and the match structure matters when
most input positions have no match.

\paragraph{Contribution and scope.}
The contribution is an explicit constructive treatment of the bitonic
objective, together with a conditional hardness reduction. The sparse
algorithm is a specialization of established dominance dynamic programming,
not a new range-searching data structure. The dense algorithm is a baseline
built from the usual longest common increasing subsequence (LCIS) scan.
Equation~\eqref{eq:peak} explains exactly which LCIS information must be
retained: an endpoint value for every candidate common peak, rather than
only a global optimum or one optimal increasing subsequence. Any LCIS
algorithm exposing this interface can be substituted; we do not claim that
existing LCIS algorithms are intrinsically unable to provide it.

Beyond this formulation, a padding construction gives an exact relation
between LCIS and LCBS optimum values. Consequently, an $O(N^{2-\varepsilon})$
algorithm for general length-$N$ LCBS inputs would contradict the known
SETH-based LCIS lower bound~\cite{duraj2019tight}. This makes precise the
worst-case limitation under which match-sensitive algorithms remain useful.
The accompanying code independently validates the formulation and illustrates
the crossover between the dense and sparse implementations.

\paragraph{Relation to the preprint.}
An earlier version of this work is available as
arXiv:2511.08958v2~\cite{rahat2026lcbs}. That version already contains the
two-pass dense and dominance-based algorithms. The present treatment
consolidates their proofs and positioning, includes the formal padding
reduction, and supplies a newly implemented and measured reproducibility
package. The two algorithms are not claimed as additions relative to that
preprint.

\section{Related work and the endpoint interface}\label{sec:related}
Classical longest common subsequence algorithms exploit either a dense
position grid or the set of equality matches. Hunt and
Szymanski~\cite{hunt1977fast} established the importance of the latter
parameterization. The value constraints in LCIS and LCBS require information
beyond the two position orders, but do not make the sparse viewpoint new.

Yang, Huang, and Chao~\cite{yang2005fast} give a quadratic LCIS algorithm.
Chan et al.~\cite{chan2007efficient} explicitly exploit the number of
matching pairs. For two length-$N$ sequences, their LCIS bound is
$O(\min\{M\log\ell,N\ell+M\}\log\log N+\operatorname{Sort}(N))$,
where $\ell$ is the LCIS length, with logarithms conventionally bounded
below by a constant. Kutz et al.~\cite{kutz2011faster} give an
output-dependent expected bound in terms of LCIS length and alphabet size.
These results are essential comparisons: match sensitivity and sparse LCIS
processing predate the present formulation. Our $M\lg_+^2M$ bound is not
an improvement over the best LCIS bounds, and we do not claim it is the
best achievable LCBS bound after adapting those algorithms.

\begin{proposition}[Endpoint lifting]\label{prop:lifting}
Suppose an LCIS procedure reports the increasing endpoint length at every match
of two sequences, with a predecessor match witnessing each nontrivial value.
Two calls to that procedure, one on the original inputs and one on the
reversed inputs, followed by $O(M)$ work, solve LCBS and reconstruct an
optimal subsequence. If $T$ and $W$ bound the time and workspace of each
call, including endpoint reporting, the combined computation takes
$O(T+S+M)$ time and $O(W+S+M)$ space.
\end{proposition}
\begin{proof}
Reversal maps a decreasing suffix at $(i,j)$ to an increasing prefix at
$(n+1-i,m+1-j)$ without changing values. Retain the first call's $O(M)$
endpoint records while making the second call. Select a peak by
\eqref{eq:peak}, and concatenate its two witnesses after removing one copy
of the peak. The validity of this concatenation is proved in
Lemma~\ref{lem:peak} below. Reversal, alignment by counting sort on match
coordinates, scanning, and reconstruction cost $O(S+M)$ additional time
and space.
\end{proof}

Proposition~\ref{prop:lifting} is a two-pass reuse of LCIS with an explicit
output requirement. A global LCIS length or one optimal witness does not
supply the required per-match scores and immutable predecessor witnesses.
Importing a faster LCIS bound requires accounting for these outputs in
both calls; adaptive parameters such as the LCIS length may differ after
reversal. Our direct dominance implementation makes these costs explicit.

For example, if both inputs equal
$\langle9,8,7,6,1,2,3,4\rangle$, their LCIS and longest common decreasing
subsequence lengths are four and five. Adding these global lengths and
subtracting one gives eight, whereas LCBS has length five. The independently
optimal halves cannot be joined at a common peak. Endpoint compatibility is
therefore necessary even though its implementation uses familiar tools.

Orthogonal range maxima and coordinate compression are standard
techniques~\cite{deberg2008computational}. The implementation below uses
nested Fenwick trees, extending the prefix-aggregation
idea~\cite{fenwick1994new} to maximum values with witnesses. No improvement
to these data structures is asserted. On the complexity side,
Duraj, K\"unnemann, and Polak~\cite{duraj2019tight} rule out strongly
subquadratic LCIS algorithms under SETH. This is compatible with logarithmic
savings in the exponent-two regime, such as the LCIS algorithms of
Duraj~\cite{duraj2020subquadratic} and Agrawal and
Gawrychowski~\cite{agrawal2020faster}; strongly subquadratic means
$O(N^{2-\varepsilon})$ for a fixed $\varepsilon>0$.

\section{Match states and peak decomposition}\label{sec:states}
A match $v=(i,j)$ has value $a(v)=A[i]=B[j]$. Write $u\prec v$ if
both coordinates of $u$ are strictly smaller than the corresponding
coordinates of $v$. A chain under $\prec$ is an embedded common subsequence.
The endpoint lengths satisfy
\begin{align}
 I(v)&=1+\max\{I(u):u\prec v,\ a(u)<a(v)\},\label{eq:inc}\\
 D(v)&=1+\max\{D(w):v\prec w,\ a(w)<a(v)\},\label{eq:dec}
\end{align}
where maxima over empty sets are zero. Removing the last element of an
increasing chain, or the first element of a decreasing chain, proves these
recurrences in both directions. A witness attaining the maximum is retained
as a predecessor or successor pointer; a singleton has a null pointer.

Figure~\ref{fig:match-grid} illustrates these states for
$A=\langle2,1,3,4,6,5,4\rangle$ and $B=\langle1,2,3,5,6,4\rangle$.
The highlighted witness $\langle1,3,5,4\rangle$ joins an increasing prefix
of length three to a decreasing suffix of length two at the same match
$v=(6,4)$, giving $I(v)+D(v)-1=4$. The arrows move forward in both input
orders even when the selected values decrease after the peak.

\begin{figure}[tbp]
\centering
\input{figures/match-grid.tex}
\caption{Match-grid view of an LCBS. Columns list $B$ and rows list $A$,
with row indices increasing downward. Every node is an equality match.
Solid arrows trace the increasing prefix; the dashed arrow traces the
decreasing suffix. The diamond marks their shared peak, whose value is five.
Unselected matches are hollow circles.}
\label{fig:match-grid}
\end{figure}

\begin{lemma}[Common-peak decomposition]\label{lem:peak}
Equation~\eqref{eq:peak} holds, and its maximizing match together with
witnesses for \eqref{eq:inc} and \eqref{eq:dec} reconstructs an LCBS.
\end{lemma}
\begin{proof}
Fix an embedding of a nonempty bitonic common subsequence and let $v$ be
its peak match. The prefix ending at $v$ has length at most $I(v)$ and
the suffix starting at $v$ has length at most $D(v)$, so the full length
is at most $I(v)+D(v)-1$. Conversely, concatenate optimum witnesses for
these two states. Every nonpeak prefix match lies strictly before $v$ in
both inputs, and every nonpeak suffix match lies strictly after $v$.
Thus the embeddings do not conflict. The values strictly increase to
$a(v)$ and then strictly decrease, giving a feasible subsequence of the
claimed length. Singleton and monotone witnesses are included. If $V$ is
empty, only the empty subsequence is feasible.
\end{proof}

Pointers must identify a fixed match record, not a mutable column buffer.
Following predecessor pointers strictly decreases both input positions;
following successor pointers strictly increases them. Hence reconstruction
terminates, visits exactly the reported number of elements, and costs time
linear in the output length. Equal values may appear on opposite sides of
the peak, but never as adjacent steps within a strict monotone half.

\section{A constructive quadratic baseline}\label{sec:dense}
The baseline adapts the LCIS row scan to report every match state.
At the start of row $i$, $d[j]$ is the maximum increasing endpoint length
at column $j$ using rows smaller than $i$. The buffer $e[j]$ identifies
a fixed match attaining that value. Within the row, a running maximum
records the best eligible earlier column with value smaller than $A[i]$.

\begin{algorithm}[t]
\caption{Increasing endpoint scan on $(A,B)$}\label{alg:dense}
\begin{algorithmic}[1]
\State $d[1..m]\gets0$; $e[1..m]\gets\bot$; initialize empty maps $L,P$
\For{$i=1,\ldots,n$}
 \State $(b,q)\gets(0,\bot)$
 \For{$j=1,\ldots,m$}
  \If{$A[i]=B[j]$}
   \State $L(i,j)\gets b+1$; $P(i,j)\gets q$
   \If{$b+1>d[j]$}
    \State $d[j]\gets b+1$; $e[j]\gets(i,j)$
   \EndIf
  \ElsIf{$B[j]<A[i]$ and $d[j]>b$}
   \State $(b,q)\gets(d[j],e[j])$
  \EndIf
 \EndFor
\EndFor
\State \Return $(L,P)$
\end{algorithmic}
\end{algorithm}

\begin{lemma}\label{lem:dense}
Algorithm~\ref{alg:dense} reports all increasing endpoint lengths and
valid immutable predecessor matches in $O(S+nm)$ time and $O(S+M)$
auxiliary space, including storage for the inputs' reversed copies.
For positive $n,m$, its scan takes $\Theta(nm)$ time.
\end{lemma}
\begin{proof}
Before processing column $j$, $b$ is the maximum of $d[k]$ over $k<j$
with $B[k]<A[i]$. Only columns whose value equals $A[i]$ can have been
updated in the current row. They are ineligible for this running maximum,
so every eligible $d[k]$ still describes an earlier row. Thus $b+1$
is exactly \eqref{eq:inc}, and $q$ is a valid predecessor. Updating
$d[j]$ preserves its maximum-over-rows meaning. Induction on rows and
columns proves the invariant. There are $nm$ constant-time scan steps,
$m$ buffer entries, and $M$ fixed endpoint records. Match records can
be assigned array identifiers as they are encountered.
\end{proof}

Every match gets a record, even when its candidate does not improve the
column maximum. This is essential for joining at later peaks with repeated
symbols. Run the scan again on reversed inputs, map each match back to the
original coordinates, and apply Proposition~\ref{prop:lifting}. The
resulting constructive baseline has the same asymptotic bounds. Its
$O(M)$ witness storage is not a linear-space reconstruction claim.

\section{The match-sensitive algorithm}\label{sec:sparse}
The dense scan touches nonmatching cells. We instead enumerate matches
and compute \eqref{eq:inc}--\eqref{eq:dec} with dominance queries.

\subsection{Preprocessing and the range-maximum structure}
Assume comparison of two symbols and operations on an index take constant
time. Sort the $S$ input occurrences by symbol, form equality groups, and
assign ordered value ranks. Within each group, output all pairs consisting
of an $A$ occurrence and a $B$ occurrence. This costs $O(S\lg_+S+M)$
time and $O(S+M)$ space. Matches can be stored in lists indexed by their
$A$ row. All bounds below are deterministic in this comparison model.

Compress the columns occurring in a match to ranks $x(v)\in[1..J]$;
let $y(v)$ be the symbol rank. For $M>0$, $J\le M$. Gaps in value
ranks are harmless. Both order relations are preserved. A data structure
$Q$ supports insertion of a scored match $(x,y,s,v)$ and returns the
maximum score and one witness in a prefix rectangle $[1..X]\times[1..Y]$.
An empty rectangle returns $(0,\bot)$.

For completeness, an implementation with the stated bounds uses a Fenwick
tree on the $J$ column ranks. Each outer node stores a sorted list of the
value ranks of all points that can update it, and an inner Fenwick tree
of maximum-score/witness pairs. These lists are built offline from the
match coordinates. Each point appears in $O(\lg_+M)$ lists; sorting
them costs at most $O(M\lg_+^2M)$ time and their total size is
$O(M\lg_+M)$. Binary search in each list locates an update coordinate
or the last query coordinate at most $Y$. Both operations visit
$O(\lg_+M)$ outer nodes and cost $O(\lg_+M)$ per node. Only increases
to stored maxima are needed. Ties may be resolved arbitrarily.

\subsection{Two sweeps}
In the forward sweep, process rows in increasing order and query
$(x(v)-1,y(v)-1)$ for each match $v$ in the current row. Add one to
the returned maximum to obtain $I(v)$. In the backward sweep, process
rows in decreasing order and replace the column rank by
$\widehat{x}(v)=J+1-x(v)$. The query
$(\widehat{x}(v)-1,y(v)-1)$ obtains the best decreasing successor.
The value coordinate is not reversed.

\begin{algorithm}[t]
\caption{Match-sensitive LCBS}\label{alg:sparse}
\begin{algorithmic}[1]
\State Enumerate $V$ in row lists; compute $x(v),y(v)$ and $J$
\If{$V=\varnothing$}\State \Return the empty sequence\EndIf
\State Initialize empty dominance structure $Q$ on $(x(v),y(v))$
\For{rows $i$ in increasing order}
 \For{matches $v$ in row $i$}
  \State $(b,P(v))\gets Q.\Query(x(v)-1,y(v)-1)$; $I(v)\gets b+1$
 \EndFor
 \For{matches $v$ in row $i$}
  \State $Q.\Update(x(v),y(v),I(v),v)$
 \EndFor
\EndFor
\State Release $Q$; initialize empty $Q$ on $(J+1-x(v),y(v))$
\For{rows $i$ in decreasing order}
 \For{matches $v$ in row $i$}
  \State $(b,T(v))\gets Q.\Query(J-x(v),y(v)-1)$; $D(v)\gets b+1$
 \EndFor
 \For{matches $v$ in row $i$}
  \State $Q.\Update(J+1-x(v),y(v),D(v),v)$
 \EndFor
\EndFor
\State Choose $p$ maximizing $I(p)+D(p)-1$
\State Recover the prefix via $P$, the suffix via $T$, and join once at $p$
\end{algorithmic}
\end{algorithm}

\begin{theorem}\label{thm:sparse}
Algorithm~\ref{alg:sparse} returns an LCBS in
$O(S\lg_+S+M\lg_+^2M)$ time and $O(S+M\lg_+M)$ auxiliary space.
\end{theorem}
\begin{proof}
At the start of forward row $i$, $Q$ contains exactly the matches of
smaller rows, with their correct increasing scores. The strict column
and value bounds select precisely the remaining conditions in
\eqref{eq:inc}. Querying the whole row before inserting it preserves
the invariant. By induction, all $I$ scores and predecessor witnesses
are correct. In the backward sweep, stored matches have larger row
indices. Mirroring makes a smaller query column equivalent to a larger
original column. The unchanged strict value bound selects exactly the
successors in \eqref{eq:dec}. Thus all $D$ scores and successor witnesses
are correct. Lemma~\ref{lem:peak} proves the returned result optimal.

Preprocessing has the cost stated above. Offline structure construction
and the two sweeps together take $O(M\lg_+^2M)$ time. The final scan
and reconstruction take $O(M)$ time. Row lists, input ranks, and match
records need $O(S+M)$ space; one dominance structure needs
$O(M\lg_+M)$ space and can be released between sweeps. The $M=0$
case returns immediately after preprocessing and satisfies the same bound.
\end{proof}

\paragraph{Strictness and immediate updates.}
Row batching makes the index invariant explicit. Immediate updates within
a row are also correct here: all matches in that row have the same value
rank, and the query stops at $y(v)-1$, excluding every such update.
This applies to both sweeps. It is a property of strict monotonicity;
replacing the query bound by $y(v)$ is incorrect even with batching,
because equal-valued matches in different rows could then extend each
other. For example, two identical constant sequences have LCBS length one.
The supplied diagnostic implementation checks immediate and batched updates
against the same independent oracle.

\paragraph{When does the bound help?}
The sparse bound is asymptotically smaller than the dense bound when
$S\lg_+S+M\lg_+^2M=o(nm)$. The sorting term must be included,
particularly for unbalanced lengths. Merely assuming $M\ll nm$ is not
sufficient because each match incurs a logarithmic overhead. For dense
inputs the simple scan can be faster and uses less auxiliary space. Neither
the theorem nor the experiments assert universal superiority over the
dense baseline or over potential adaptations of faster LCIS algorithms.

\section{A conditional worst-case barrier}\label{sec:hardness}
The inclusion of increasing subsequences among bitonic ones does not alone
reduce LCIS optimization to LCBS: a decreasing phase might improve the
optimum. A fresh increasing suffix eliminates that possibility in an optimal
solution.

\begin{lemma}[Padding identity]\label{lem:padding}
Let $Z=\langle z_1,\ldots,z_K\rangle$ be strictly increasing, with all
its symbols fresh and greater than every symbol occurring in $A$ or $B$.
For $K\ge1$,
\begin{equation}\label{eq:padding}
 \LCBS(A\circ Z,B\circ Z)
 =\max\{\LCBS(A,B),\LCIS(A,B)+K\}.
\end{equation}
\end{lemma}
\begin{proof}
A common subsequence using no fresh symbol has length at most
$\LCBS(A,B)$ when bitonic. A bitonic common subsequence using a fresh
symbol must be purely increasing: original symbols precede all fresh
symbols in both inputs, the transition to the fresh suffix is increasing,
and fresh symbols appear in increasing order. A prior decrease would
therefore be followed by an increase, while a later decrease is impossible.
Its original part has length at most $\LCIS(A,B)$ and its fresh part
has length at most $K$. Conversely, an original LCBS remains feasible,
and appending all of $Z$ to an LCIS gives a strictly increasing common
subsequence of length $\LCIS(A,B)+K$. These two constructions achieve
both alternatives in the maximum.
\end{proof}

Choose $K=\min\{n,m\}+1$. Since $\LCBS(A,B)\le\min\{n,m\}$,
the second alternative strictly dominates, and
\begin{equation}\label{eq:reduction}
 \LCBS(A\circ Z,B\circ Z)=\LCIS(A,B)+K.
\end{equation}
The construction has linear size. Fresh symbols can be represented as
tagged ordered symbols following the original alphabet. For integer inputs,
a linear scan finds the maximum and the fresh symbols can be consecutive
larger integers. Every optimal padded solution is increasing and contains
all fresh symbols; otherwise a missing fresh symbol could be inserted.
Removing the suffix therefore recovers an LCIS witness as well as its
length. The construction adds exactly $K$ matches, since each new symbol
occurs once in each input and matches no old symbol.

\begin{corollary}\label{cor:seth}
Assuming SETH, LCBS on two length-$N$ integer sequences has no
$O(N^{2-\varepsilon})$ algorithm for any fixed $\varepsilon>0$.
\end{corollary}
\begin{proof}
Given two length-$N$ LCIS inputs, take $K=N+1$. Each padded input has
length $2N+1$. A strongly subquadratic LCBS algorithm, followed by
subtracting $K$, would solve LCIS in strongly subquadratic time,
contradicting the lower bound of Duraj et al.~\cite{duraj2019tight}.
\end{proof}

This is a worst-case, conditional statement for unrestricted inputs. It
does not establish an unconditional quadratic lower bound, optimality of
the logarithmic factors in Theorem~\ref{thm:sparse}, or hardness of every
sparse regime. In particular, a bound depending on $M$ is consistent with
the corollary because $M$ can be quadratic in $N$.

\section{Reproducible implementation and experiments}\label{sec:experiments}
The accompanying supplement contains new Python implementations of the dense
scan and the compressed two-dimensional Fenwick algorithm. Both recover
match-coordinate witnesses and include preprocessing and reconstruction in
their measured runtime. A third routine uses immediate strict updates as a
diagnostic. No third-party Python packages are needed to run the algorithms,
tests, or benchmarks. The measurements below were generated specifically
for this revision; they do not reuse timings from earlier drafts.

\subsection{Independent correctness checks}
An independent oracle enumerates all subsequences of the first input and
tests their bitonicity and membership in the second input. Exhaustive
testing covers all pairs of sequences of length zero through four over a
three-symbol alphabet: $121^2=14\,641$ pairs. Another 1,000 seeded random
pairs have lengths zero through eight and values from $-3$ through $4$.
For every pair, the tests compare the dense, batched sparse, and immediate
sparse optimum lengths with the oracle, and check all reconstructed
coordinates and values. They additionally compare every forward and
backward endpoint score between the dense and sparse implementations and
check the LCIS padding equality. All 15,641 test cases (14,641 exhaustive
and 1,000 randomized) passed these checks.
The validation log and test seeds are included. These finite tests check
the implementation; the preceding proofs establish general correctness.

\subsection{Measurement protocol}
Inputs are independently uniform integer sequences over
$\{0,\ldots,\sigma-1\}$, with $n=m=N$. Thus $\mathbb{E}[M]=N^2/\sigma$.
The density series fixes $N=600$ and varies the alphabet size; the scaling
series sets $\sigma=N$. Each configuration uses three independently seeded
input pairs and three timed repetitions per pair and algorithm. Each
algorithm is warmed up once before the series, and execution order
alternates. Garbage collection runs before each timed call and remains
enabled. Input generation and validation are outside the timed region.
The clock is Python's \texttt{perf\_counter\_ns}.

For each algorithm we first take the median of three repetitions on each
input pair, then the median of those three medians. Table~\ref{tab:results}
reports these values, the median match count, and the ratio of aggregate
dense time to aggregate sparse time. The ratio is calculated before
rounding; it is not the median of individual speedups. All 180 raw timings,
input-generation seeds, and the execution environment are included in the
supplement. Each measured result was validated, and the two implementations
agreed on optimum length for every benchmark input.

\input{environment.tex}

\begin{table}[t]
\centering
\caption{Fresh end-to-end measurements. Ratio is dense time divided by
sparse time; values below one favor the dense implementation.}
\label{tab:results}
\small\input{results-table.tex}
\end{table}

\subsection{Observed tradeoff and limitations}
Figure~\ref{fig:runtime-crossover} plots the fixed-$N$ density measurements
from Table~\ref{tab:results} against their median match counts. The ordering
of the measured runtimes switches between the tested configurations with
$M=1425$ and $M=5597$; the shaded interval indicates this observation,
not an estimated exact threshold. Both axes are logarithmic, and the line
segments only connect measurements.

\begin{figure}[tbp]
\centering
\input{figures/runtime-plot.tex}
\caption{Runtime comparison on the six density-series configurations with
$n=m=600$. Each point uses the same aggregate runtime and median match
count as Table~\ref{tab:results}. Dashed squares denote the dense baseline;
solid circles denote the match-sensitive implementation. The shaded band
lies between adjacent tested configurations where the faster method
changes. No measurements within that band or universal crossover are claimed.}
\label{fig:runtime-crossover}
\end{figure}

At $N=600$, the dense routine is faster for alphabet sizes 4, 16, and 64.
The sparse routine becomes faster at the tested alphabet size 256, and its
advantage grows as matches become rarer. The measured ratio ranges from
approximately 0.21 at $\sigma=4$ to 26.32 at $\sigma=4096$. In the
$\sigma=N$ series it increases from 2.39 at $N=200$ to 11.19 at $N=1600$.
These observations illustrate the cost of managing matches and the benefit
of avoiding a full grid scan. They do not locate a universal crossover.

We have not implemented and validated LCBS adaptations of the adaptive
LCIS methods of Chan et al.~\cite{chan2007efficient} and
Kutz et al.~\cite{kutz2011faster}, so the experiments compare only our two
reference implementations. A like-for-like comparison through
Proposition~\ref{prop:lifting} must include both passes, per-match outputs,
and reconstruction; timing an LCIS-only solver would measure a different
task. This omission establishes no barrier to adaptation, and the results
imply no superiority over those methods. The study also excludes optimized
native implementations and peak-memory measurements; its single machine,
synthetic inputs, and small seed sample do not support broad application
claims. Python allocation, dictionary lookup, cache effects, and system
load influence absolute times. Earlier application-oriented measurements
are not used because their original scripts and logs were unavailable.

\section{Conclusion}
LCBS can be solved constructively by joining compatible increasing and
decreasing endpoint states at a common match. Reporting these states is the
precise interface needed to reuse LCIS techniques. A standard dominance
implementation yields a transparent match-sensitive bound, while a simple
padding identity transfers the SETH-based worst-case barrier from LCIS.
The supplied implementation and fresh measurements document both sparse
gains and dense overhead. Useful next questions are whether faster LCIS
methods preserve all needed endpoint witnesses at their advertised cost,
and how much reconstruction space can be saved without recomputing a
large fraction of the states.

\section*{Acknowledgements and declarations}
The authors acknowledge the American International University-Bangladesh
for support. No funding was received for this study. The authors declare
no competing interests.

\paragraph{Data and code availability.}
The submission supplement contains the complete reference implementations,
independent oracle, validation report, seeded input generator, raw timing
records, environment metadata, and commands to reproduce the results. All
experimental inputs are synthetic and can be regenerated from the recorded
seeds. No external dataset is required.

\paragraph{Author contributions.}
Md. Tanzeem Rahat developed the formulation and wrote the manuscript.
Md. Manzurul Hasan supervised the work and revised the manuscript.
Sadia Sharmin contributed to supervision and reviewed the results.

\bibliographystyle{fundam}
\bibliography{references}
\end{document}

%% file: figures/match-grid.tex
% Editable vector diagram. Coordinates are (column j, 8-row i).
\begin{tikzpicture}[
  x=0.87cm,y=0.55cm,font=\small,
  match/.style={circle,draw=black!65,fill=white,minimum size=5pt,inner sep=0pt},
  selected/.style={circle,draw=blue!55!black,fill=blue!55!black,minimum size=5pt,inner sep=0pt},
  peak/.style={diamond,draw=orange!70!black,fill=orange!35,minimum size=10pt,inner sep=1pt},
  rising/.style={-{Latex[length=2mm]},blue!55!black,very thick},
  falling/.style={-{Latex[length=2mm]},orange!75!black,very thick,dashed}
]
\foreach \x in {1,...,6} {\draw[black!14] (\x,0.8)--(\x,7.2);}
\foreach \y in {1,...,7} {\draw[black!14] (0.8,\y)--(6.2,\y);}
\node at (-1.1,7.7) {$i$};
\node at (-0.25,7.7) {$A[i]$};
\node at (-0.25,9.35) {$j$};
\node at (-0.25,8.55) {$B[j]$};
\foreach \j/\b in {1/1,2/2,3/3,4/5,5/6,6/4} {
  \node[font=\scriptsize,text=black!65] at (\j,9.35) {\j};
  \node at (\j,8.55) {$\b$};
}
\foreach \i/\a in {1/2,2/1,3/3,4/4,5/6,6/5,7/4} {
  \node[font=\scriptsize,text=black!65] at (-1.1,{8-\i}) {\i};
  \node at (-0.25,{8-\i}) {$\a$};
}
% All seven matching position pairs, with the selected path overlaid.
\foreach \x/\y in {2/7,1/6,3/5,6/4,5/3,4/2,6/1} {
  \node[match] at (\x,\y) {};
}
\draw[rising] (1,6)--(3,5);
\draw[rising] (3,5)--(4,2);
\draw[falling] (4,2)--(6,1);
\node[selected] at (1,6) {};
\node[selected] at (3,5) {};
\node[peak] at (4,2) {};
\node[selected] at (6,1) {};
\node[above right=2pt,font=\scriptsize] at (4,2) {$v=(6,4)$};
\node[anchor=west,align=left] at (7.0,7.0) {\textbf{One optimal witness}\\[3pt]
  $\langle1,3,\underbrace{5}_{\text{peak}},4\rangle$};
\draw[rising] (7.0,5.1)--(7.8,5.1);
\node[anchor=west] at (8.0,5.1) {Increasing prefix};
\draw[falling] (7.0,4.35)--(7.8,4.35);
\node[anchor=west] at (8.0,4.35) {Decreasing suffix};
\node[match] at (7.2,3.6) {};
\node[anchor=west] at (7.7,3.6) {Other match};
\node[anchor=west,align=left] at (7.0,1.9) {$I(v)=3,\quad D(v)=2$\\[4pt]
  $I(v)+D(v)-1=4$};
\node[anchor=west,font=\scriptsize,text=black!70] at (7.0,0.6) {The shared peak is counted once.};
\end{tikzpicture}

%% file: environment.tex
Measurements used CPython 3.12.14 on Windows 11 (build 26200), with an AMD Ryzen 7 5700X processor (8 cores, 16 logical processors) and approximately 32 GiB RAM. The algorithms execute sequentially in one Python process.

%% file: results-table.tex
\begin{tabular}{lrrrrrr}
\hline
Series & $N$ & $\sigma$ & $M$ & Dense (s) & Sparse (s) & Ratio\\
\hline
Density & 600 & 4 & 89895 & 0.1453 & 0.6788 & 0.21\\
Density & 600 & 16 & 22410 & 0.0614 & 0.2016 & 0.30\\
Density & 600 & 64 & 5597 & 0.0440 & 0.0586 & 0.75\\
Density & 600 & 256 & 1425 & 0.0376 & 0.0178 & 2.11\\
Density & 600 & 1024 & 340 & 0.0371 & 0.0042 & 8.84\\
Density & 600 & 4096 & 86 & 0.0371 & 0.0014 & 26.32\\
Scaling & 200 & 200 & 200 & 0.0040 & 0.0017 & 2.39\\
Scaling & 400 & 400 & 429 & 0.0174 & 0.0052 & 3.37\\
Scaling & 800 & 800 & 814 & 0.0685 & 0.0104 & 6.56\\
Scaling & 1600 & 1600 & 1570 & 0.2688 & 0.0240 & 11.19\\
\hline
\end{tabular}

%% file: figures/runtime-plot.tex
% Vector plot of the six measured fixed-N density configurations.
\begin{tikzpicture}
\begin{loglogaxis}[
  width=12.8cm,height=6.8cm,
  xlabel={Median number of matches $M$},
  ylabel={Median runtime (seconds)},
  xmin=60,xmax=140000,ymin=0.001,ymax=1.2,
  xtick={100,1000,10000,100000},
  ytick={0.001,0.01,0.1,1},
  yticklabels={$0.001$,$0.01$,$0.1$,$1$},
  grid=major,grid style={black!15},
  tick label style={font=\small},label style={font=\small},
  legend style={font=\small,draw=none,fill=white,at={(0.03,0.97)},anchor=north west},
  legend cell align=left,
  clip=true
]
\path[fill=black!7] (axis cs:1425,0.001) rectangle (axis cs:5597,1.2);
\addplot[black,thick,dashed,mark=square*,mark size=2.4pt]
  table[x=matches,y=dense_seconds,col sep=comma]{figures/runtime-data.csv};
\addlegendentry{Dense}
\addplot[blue!55!black,thick,mark=*,mark size=2.6pt]
  table[x=matches,y=sparse_seconds,col sep=comma]{figures/runtime-data.csv};
\addlegendentry{Match-sensitive}
\node[anchor=north,font=\scriptsize,align=center,fill=white,inner sep=2pt]
  at (axis cs:2824,1.05) {Observed\\switch interval};
\node[anchor=south east,font=\small,fill=white,inner sep=2pt]
  at (rel axis cs:0.98,0.03) {$n=m=600$};
\end{loglogaxis}
\end{tikzpicture}